\documentclass[10pt]{article}
\usepackage{amsmath}
\usepackage{graphicx}
\usepackage{subfig}

\include{Qcircuit}

\begin{document}

\title{Reliable quantum circuits have defects}
\date{}
\author{Alexandru Paler, Austin G. Fowler, Robert Wille}
\maketitle

State of the art quantum computing architectures are founded on the decision to use scalable but faulty quantum hardware in conjunction with an efficient error correcting code capable of tolerating high error rates. The promised effect of this decision is that the first large-scale practical quantum computer is within reach. Coming to grips with the strategy and the challenges of preparing reliable executions of an arbitrary quantum computation is not difficult. Moreover, the article explains why defects are good.

\section{``It can only be attributable to human error''}

In a not too distant future, a quantum computer engineer will be confronted with the problem of automating the compilation of what a user wishes to execute (quantum algorithm) to instructions which a quantum computer is able to execute. 

A quantum algorithm is implemented as a quantum circuit formed of quantum gates operating on quantum bits (qubits). Executing a quantum circuit is different compared to classical circuit execution. The wires and the gates of a classical circuit are implemented in hardware. In contrast, for a quantum circuit, only the qubits may be seen as part of the hardware, because the gates are understood as instructions for transforming qubit states. There are multiple related gate sets which can be used to express quantum algorithms, and their relation is similar to how classical programming languages are compiled from high level ones into lower level and finally assembler instructions.

For solving the problem, the quantum computer engineer should learn about how defects are useful for constructing reliable quantum circuits \cite{raussendorf2007fault,fowler2012surface}. He needs to devise a method to automatically transform the high level description of the quantum algorithm into an equivalent low level description. This would be straightforward in an ideal world, but in reality quantum hardware is faulty. The solution is to chose a suitable quantum error correcting code and to compile the algorithm into an intermediate description language which guarantees very high computational reliability. The surface quantum error correcting code is chosen due to its excellent error correcting properties and very low resource overheads. The surface code can be used for hardware failing less than 1\% of time.

It is reasonable to assume that the first quantum computer will be built from faulty hardware entities arranged in a two dimensional lattice (e.g. Figure~\ref{fig:defects}). Each entity in the lattice represents a physical qubit that can be manipulated individually or interacted with its nearest neighbours. Physical qubits can be either on (when actively manipulated) or off (when not being used). The available computational resources are constraint by the lattice area (number of physical qubits) and time (number of interaction rounds). 

The engineer faces the challenge of compiling (as efficiently as possible with respect to the computational resources) an algorithm into surface code elements. This article will alleviate the engineer's fear of defects, because these are basic error corrected elements used by the surface code.

\subsection{Quantum Circuit Ingredients}

Quantum circuits have their own particularities, given that they describe computations based on quantum physical effects. Firstly, a quantum circuit has the same number of inputs and outputs. Secondly, all the gates have the same number of inputs and outputs, and thirdly, the state of arbitrary qubits cannot be copied. From a diagrammatic point of view, a quantum circuit is a set of horizontal wires interrupted by quantum gates (e.g. Figure~\ref{fig:circuit}). Quantum circuit wires are qubit abstractions. The state of a qubit is transformed by each gate applied to that wire after a left-to-right traversal. Circuit inputs are on the left side, the outputs on the right side.

\begin{figure}
\includegraphics[width=.9\columnwidth]{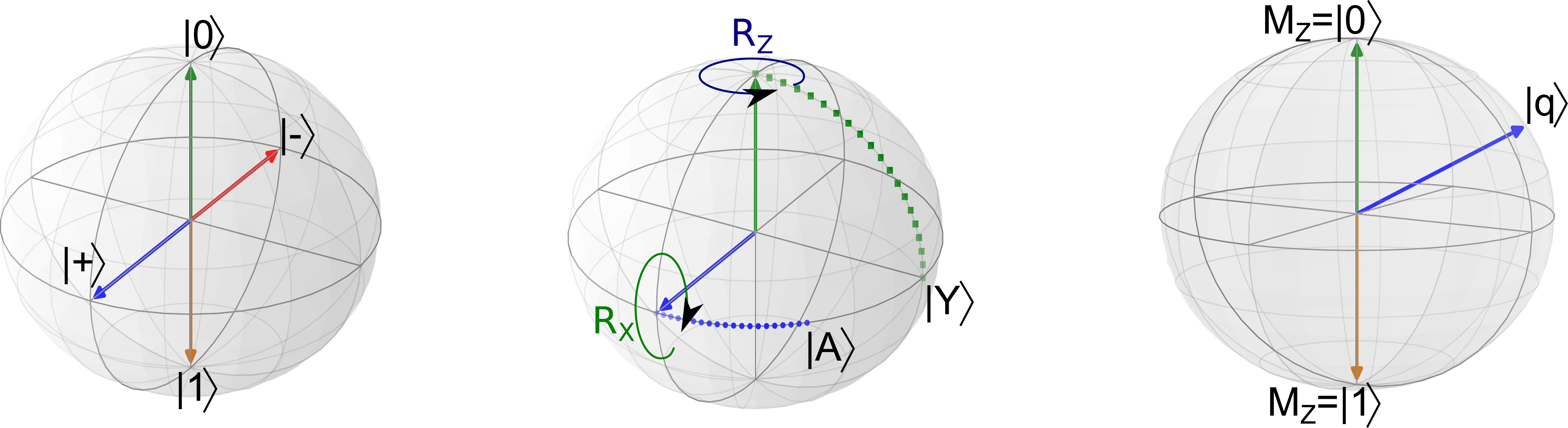}
\caption{Visualisation of state initialisation, gate applications and measurement: left) A qubit can take any state between north pole ($\ket{0}$) and south pole ($\ket{1}$). The poles are called computational basis states. A qubit's state can also be on the equator in $\ket{+}$ or $\ket{-}$.; middle) Single qubit gates are rotations around an axis. The bit flip transforming $\ket{0}$ into $\ket{1}$ is a $\pi$ rotation around the X axis. Similarly, the $\ket{+}$ is transformed into $\ket{A}$ by a $\pi/4$ rotation around the Z axis; right) A measurement returns one of the two possible states, where the axis touches the unit sphere surface; for example, a Z basis measurement (symbolised by $M_Z$) returns either $\ket{0}$ or $\ket{1}$.}
\label{fig:spheres}
\end{figure}

It would be difficult to discuss circuits and computations without introducing a few technical details. The state of a qubit named $q$ is a two dimensional complex vector denoted $\ket{q}$ and imagined to indicate a point on the surface of a three dimensional unit sphere (Figure~\ref{fig:spheres}). Quantum gates rotate qubit states around a sphere axis and each single qubit gate can be decomposed into three rotations around two orthogonal axes, for example Z and X. Considering the arbitrary quantum gate $G$, we will write $G=R_Z(\alpha)R_X(\beta)R_Z(\gamma)$, where $R_X$ and $R_Z$ are rotation operators around the X and Z axes, and $\alpha, \beta, \gamma$ are rotation angles. This decomposition is a first example of how a high level description (gate $G$) is compiled into a lower level description (only rotation gates). Single qubit measurements are probabilistic and performed around sphere axis. The probability of the measurement result depends on the angle between the measured state and the measurement axis. Multi-qubit gates exist too: for example, the CNOT gate is one possibility to create the quantum specific phenomenon of entanglement. Quantum entanglement occurs when pairs of qubits are interacted in a manner such that the state of each qubit cannot be described independently. The CNOT gate performs a bit flip of a computational state (target) if another state is $\ket{1}$ (control).

\subsection{Defects, Braids and Distillations}

Bad environment has also an influence on quantum circuits, resulting in faulty qubit initialisations and measurements or faulty quantum gate applications. The majority of the environment-induced faults can be mitigated by quantum error correcting codes. From the perspective of the surface code, logical qubits are encoded and operated (initialisations, measurements and gates) by switching off sets of physical qubits in the lattice and interacting only the qubits still on. It suffices to mention that there are two methods (primal and dual) for manipulating single physical qubits and for saving the details about how the qubits interact with each other. Physical qubits, depending on their manipulation method, are switched on and off in turns: primal, dual, primal \ldots

\begin{figure}
\centering
\subfloat[]{
  \label{fig:defects}
  \includegraphics[width=.9\columnwidth]{figs/defects.pdf}
}
\\
\subfloat[]{
  \label{fig:braidcnot}
  \includegraphics[width=0.7\columnwidth]{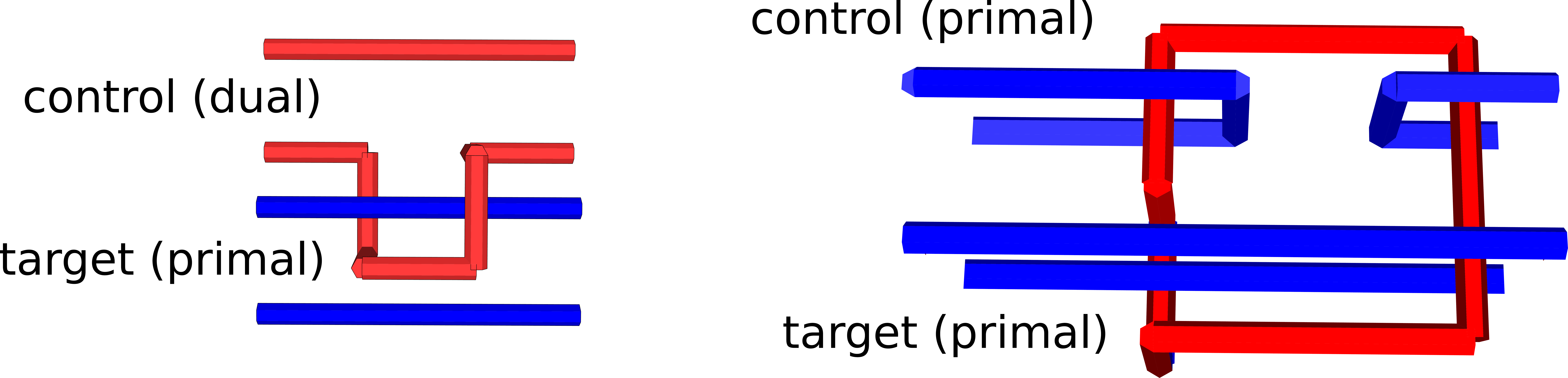}
}
\caption{From physical qubits to defects and braids. b) A lattice of 16 physical qubits which can be switched off (green) is used in rounds by switching on primal qubits (blue) and dual qubits (red); a primal defect (blue) abstracts the set of switched off primal qubits (e.g. one qubit in two rounds); a dual defect (red) abstracts switched off dual qubits (e.g. one qubit in two rounds); b) The dual-primal CNOT is a single braid, and a primal-primal CNOT is obtained by using three braids.}
\end{figure}

A defect abstracts how a set of switched off lattice physical qubits is evolving in time (Figure~\ref{fig:defects}), and depending on the physical qubits type, the defect can be either primal or dual.  A logical qubits is formed by pairs of same type defects and, as a result, the surface code allows the construction of primal and dual logical qubits.

A logical CNOT gate (Figure~\ref{fig:braidcnot}) is a braid between defects of opposite type (a primal and a dual): the dual logical qubit controls the primal logical qubit (target). Braids between defects of the same type leave the corresponding logical qubit states untransformed; the result is a logical identity gate. Braiding is the only straightforward operation between defects, implying that arbitrary quantum circuits, consisting of logical qubit initialisations, logical CNOT gates and logical qubit measurements, can be easily protected the surface code. 

Error correction capability of the surface code (its distance) is a function of defect circumference and defect distances: construct distant, thick defects when using very faulty hardware, and construct close, thin defects when the hardware is less faulty. Code distance is not discussed in the following because it does not influence the definition of defects and braids

The surface code will not solve all environment related issues. Although an initialised qubit will be protected against errors, it may have a low fidelity: using the sphere visualisation, there is a large distance between the actual state and the ideal state. Fidelity is increased by distillation procedures \cite{bravyi2005universal} expressed as subcircuits. These take multiple low fidelity instances of a state and output a single high fidelity state. Consequently, the surface code will have to protect circuits including distillation procedures.

\section{The RISC of Quantum Circuits}

Reduced instruction set computing (RISC) was proposed as a way to increase classical computing performance, but the performance of quantum computers is not a thoroughly discussed research topic for the moment. However, there are sufficient reasons why a reduced set of quantum gates is useful. On the one hand, design automation methods can focus on a common framework; on the other, there are known efficient methods to actively protect specific gates against errors.

\subsection{T is the Difficult Gate}

Quantum computers seem more powerful for particular tasks which are exponentially difficult for classical computers. There are exceptions: quantum computations using only Clifford gates; but these are not universal, cannot express the full capabilities of quantum computing. Clifford gates are the Hadamard $H=R_Z(\pi/2)R_X(\pi/2)R_Z(\pi/2)$, $V=R_X(\pi/2)$, $S=R_Z(\pi/2)$, the CNOT, and any other gate combination of the previous (e.g. $SHV$). Only the Clifford+T gate set is universal and indeed exponentially difficult for classic computers, because of the gate $T=R_Z(\pi/4)$.

All gate types are translatable into Clifford+T, and research has focused lately on this set. Computations including state distillations can be protected by the surface code if all the circuit's gates are decomposed into Clifford+T and then into ICM (single qubit initialisations, CNOT gates and single qubit measurements) \cite{paler2015compiler}, the smallest set of gates. 

\subsection{Initialise, Entangle and Measure}

\begin{figure}
\centering
\small
\subfloat[ ]{
	\label{circ:ftv}
	\Qcircuit @C=0.3em @R=.3em {
		\lstick{\ket{\psi}} & \targ & \qw  &\qw & \rstick{S\ket{\psi}}\\
		\lstick{\ket{Y} }&\ctrl{-1}&\gate{M_X}
	}
}
\hfil
\subfloat[ ]{
	\label{circ:ftt}
 \Qcircuit @C=.3em @R=.3em {
 	\lstick{\ket{\psi}} & \targ & \qw & \qw & \qw & \qw &\gate{M_Z} & \controlo \cw & \control \cw \\
	\lstick{\ket{A}} & \ctrl{-1} & \ctrl{1} & \targ & \qw & \qw & \qw &\gate{M_Z} \cwx & \gate{M_X}\cwx \\
		\lstick{\ket{0}} & \qw &\targ & \qw & \targ &\qw & \qw & \gate{M_X} \cwx&\gate{M_Z}  \cwx\\
		\lstick{\ket{Y}} & \qw &\qw & \ctrl{-2} & \qw &\ctrl{2} &\qw & \gate{M_X} \cwx&\gate{M_Z} \cwx\\
		\lstick{\ket{+}} & \qw &\qw &\qw &\ctrl{-2} & \qw& \ctrl{1} & \gate{M_Z} \cwx&\gate{M_X} \cwx\\
		\lstick{\ket{0}} & \qw &\qw & \qw & \qw & \targ & \targ & \qw & \qw & \qw & \rstick{T\ket{\psi}}
	}
}
\caption{Quantum circuits. Horizontal wires abstract the manipulation of qubits. The CNOT gate is symbolised by the horizontal line connecting the $\bullet$ (control qubit) and the $\oplus$ (target qubit). Controlled measurements, indicated by $\|$, determine the measurement basis of a qubit depending on the result of another qubit (e.g. the upper Z basis measurement). Following circuits are ICM representations: a) The S gate; b) The T gate.}
\label{fig:circuit}
\end{figure}
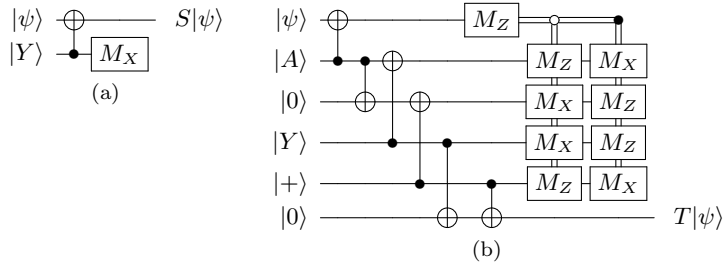

ICM are the RISC of surface code protected quantum circuits. Clifford+T gates are translated into subcircuits of only qubit initialisations, CNOT gates and qubit measurements. Initialisations can be either one of four possible states ($\ket{0}$, $\ket{+}$, $\ket{A}$, $\ket{Y}$), while measurements are of two types: single qubit independent ones (Figure~\ref{circ:ftv}), and classically controlled measurements (Figure~\ref{circ:ftt}).

An ICM single qubit rotation (e.g. V, S, T) about an angle $\theta$ is implemented by entangling an ancilla qubit (initialised into one of the four states) to the qubit representing the state to be rotated, and measuring one of the two qubits. Quantum measurements are probabilistic and will yield a rotation by $\theta$ or by $-\theta$ (a rotation about $\theta$ in the opposite direction). The latter situation requires the wrongly rotated state to be corrected by an additional rotation of $2\theta$. For the V and S gates, the correction can be tracked through the CNOT circuit and does not need to be implemented as a gate. The ICM formulation of the S gate is illustrated in Figure~\ref{circ:ftv}.

The ICM T gate ($R_Z(\pi/4)$) is slightly more complex, because it requires an S gate correction ($R_Z(2\pi/4)=R_Z(\pi/2)=S$) which cannot be tracked. It needs to be executed inside the circuit and, therefore, four other ancilla qubits are entangled and measured depending on the topmost $M_Z$ result. The circuit performs a $R_Z(\pi/4)$ rotation irrespective of the upper $M_Z$ result.

\subsection{``Forget about small efficiencies''}

Compiling a quantum circuit into surface code elements is halfway finished. The ICM form introduced a quantum circuit wire for each ancilla qubit, but not all the qubits are required simultaneously during a computation. Independent qubits can share the same wire, so that a qubit measurement is not preceded by another qubit's initialisation. This leads to less physical resources required for error correcting the computation.

\begin{figure}
\includegraphics[width=0.9\columnwidth]{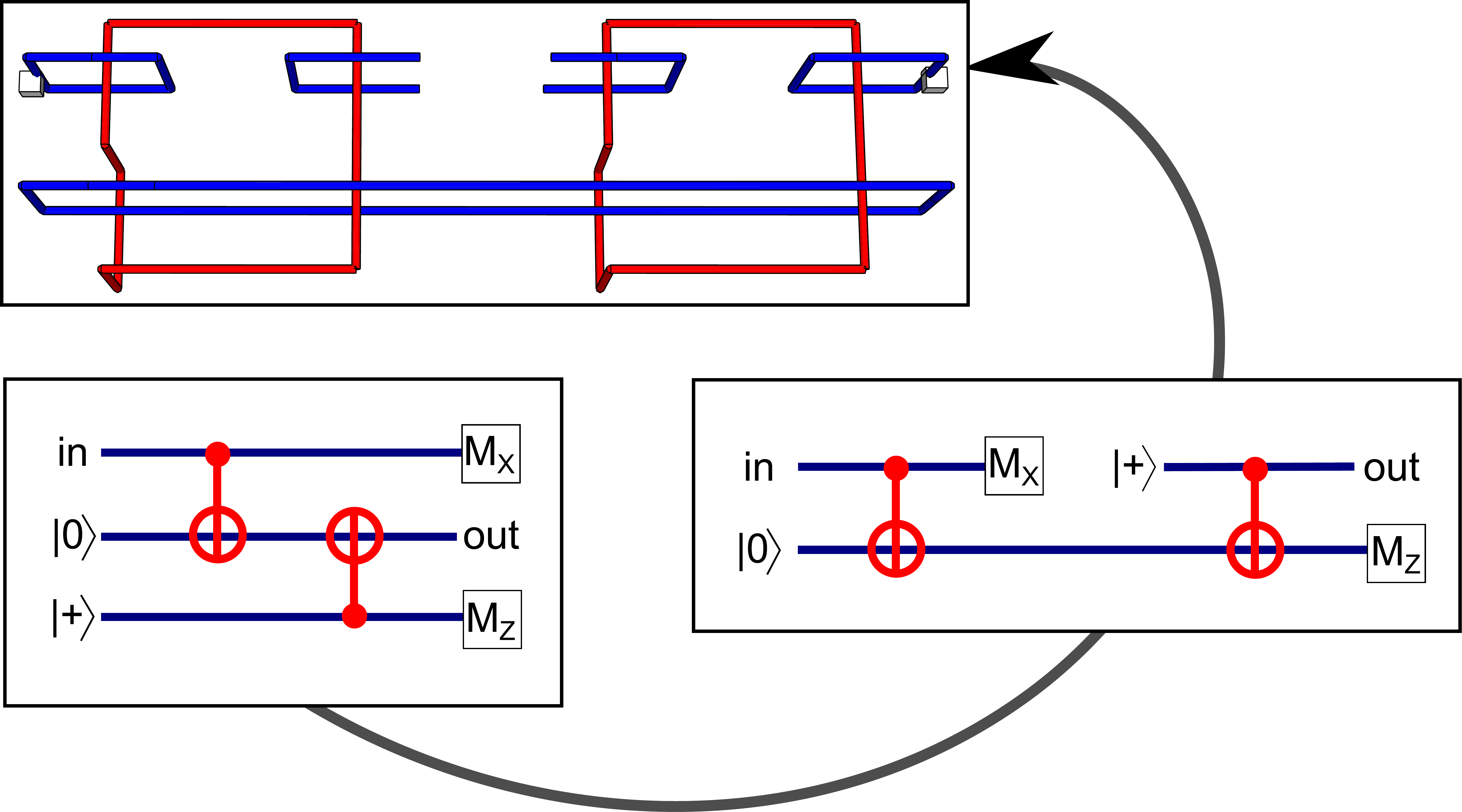}
\caption{An assembly of braided defects is obtained from an optimised circuit (two qubits share the same wire). Logical qubits are pairs of blue (primal) defects. Each CNOT from the circuit is represented by a red (dual) defect braided three times around primal defects. In the assembly, the initialisation and the measurement of a logical qubit depends on the existence of a third defect connecting defect pairs (e.g. compare $\ket{0}$ and $\ket{+}$ initialisations).}
\label{fig:rec}
\end{figure}

\section{The Defects Ensemble}

At last, the quantum computer engineer can protect a quantum circuit against the bad environment. The circuit is translated into an assembly of braided defects (e.g. Figure~\ref{fig:rec}). It is not specified if a qubit should be primal or dual, and the solution is to consider all the qubits being primal, and implement a logical CNOT between primal qubits. Each CNOT from the circuit is implemented by the primal only braided logical CNOT from Figure~\ref{fig:braidcnot}.

\begin{figure}
\includegraphics[width=.9\columnwidth]{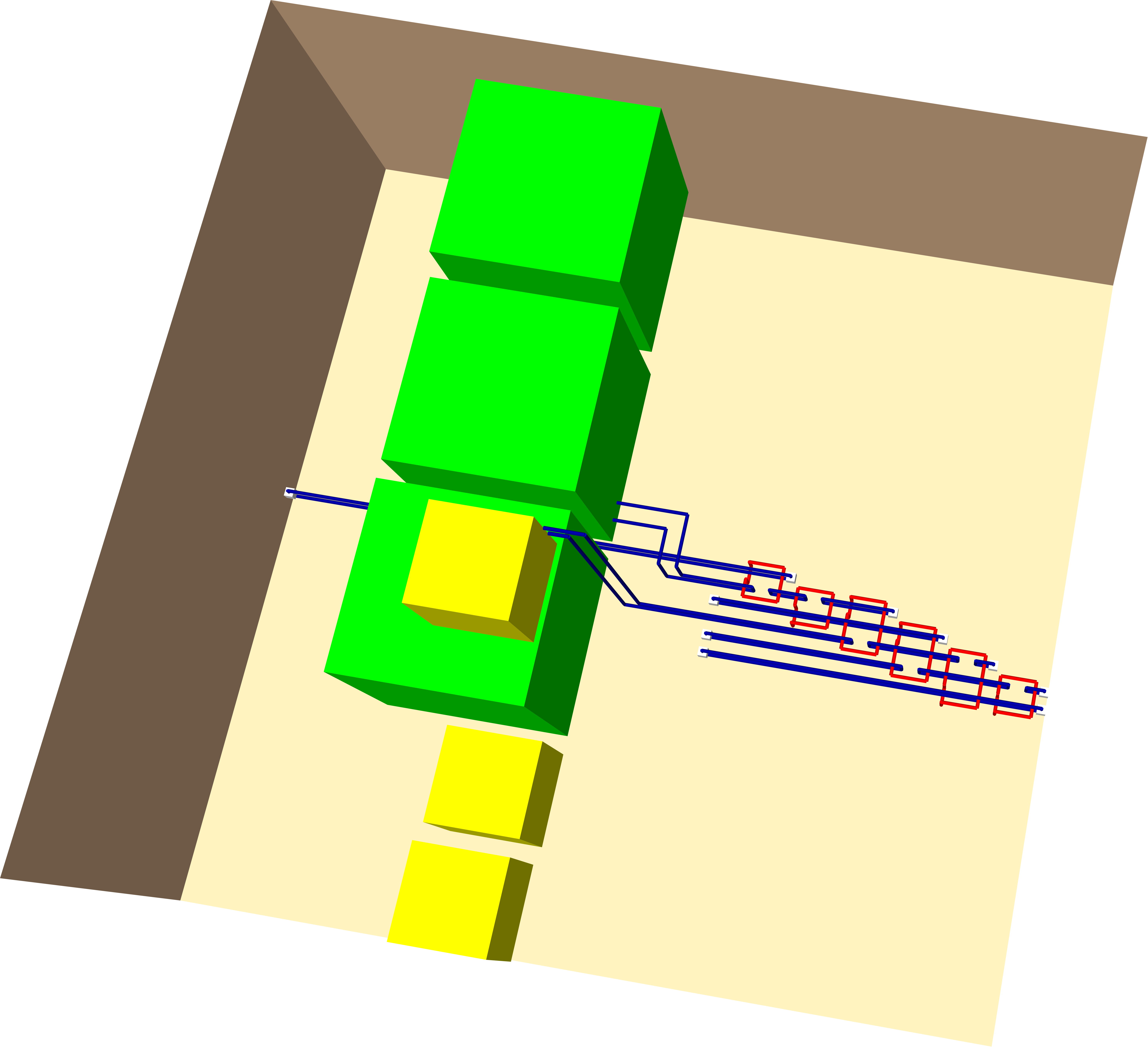}
\caption{The defect assembly corresponding to the T gate from Figure~\ref{circ:ftt} requires high fidelity $\ket{Y}$ and $\ket{A}$ states. Three distillations of each type are executed (yellow and green) and only the successful ones are connected to the circuit by pairs of defects. The six logical qubits from the ICM circuit are the six parallel pairs of blue (primal) defects. Each CNOT from the circuit corresponds to a red (dual) trice braided defect. This is a resource unoptimal placement of boxes because an imagined bounding box (orange) is largely unoccupied.}
\label{fig:synth}
\end{figure}

The resulting circuit (e.g. Figure~\ref{fig:synth}) includes distillations symbolised by boxes of different volumes ($\ket{A}$ and $\ket{Y}$). Boxes are placeholders of surface code protected versions of the distillation subcircuits. Distillations are probabilistic (may not succeed and the output state has low fidelity) and heralded (it is known if distillation succeeded). The engineer computes the number of boxes to be sufficient for very computational reliability, and lets all boxes execute in order to know which were successful and which not. Only outputs of successful boxes are usable, and each high fidelity output is connected to the circuit by a pair of defects to where it is needed.

\section{TODO: Build quantum computer}

The T gate, the one which makes quantum computations difficult, greatly influences the defects assembly too. The gate requires high fidelity $\ket{A}$ states, and the total number of corresponding distillations combined with their resource requirements is so high that it almost monopolises the cost of error correcting an arbitrary quantum computation.

Instead of a conclusion, the quantum computing engineer should not be afraid of defects and be motivated to investigate ways to optimise distillations and their placement in assemblies. The engineer can learn more about this topic by using the complete introduction to the surface code offered in \cite{fowler2012surface} and the description of how defect assemblies are generated \cite{paler2016synthesis}.

\bibliographystyle{plain}
\bibliography{xrds}

\end{document}